\documentstyle[12pt]{article}
\begin{document}

\pagestyle{empty}
\renewcommand{\baselinestretch}{1.1}
\parskip4pt
\setlength{\textwidth}{16cm}
\setlength{\textheight}{22cm}
\addtolength{\oddsidemargin}{-1.5cm}
\addtolength{\topmargin}{-1cm}

\hoffset1truecm

\tolerance 1000
\overfullrule0mm

\def\del{\Delta m_{\nu}^2\over2}
\def\A{A^{\star}}

\topskip2mm
\begin{titlepage}

\begin{center}


{\large\bf
NEUTRINO SPIN AND FLAVOUR CONVERSION AND  OSCILLATIONS
 IN  MAGNETIC FIELD}\\


\vspace{2.5truecm}
{\large A.M. Egorov}\\

\vspace{2truemm}

{\sl Department of Quantum Statistic Physics, Physics Faculty,
 Moscow State
University, 119899 Moscow, Russian Federation}

\vspace{2truemm}
{\large G.G. Likhachev}\\

\vspace{2truemm}
{\sl
Department of General Physics and High Mathematics,
 State
University on Eartharrangment, Kazakov Street, 15, Moscow, Russian
Federation}

\vskip2truemm
{\large A.I. Studenikin}\footnote{\normalsize E-mail:
studenik@srdlan.npi.msu.su}\\

\vskip2truemm
{\sl Department of Theoretical Physics, Physics Faculty,\\
 Moscow State
University, 119899 Moscow, Russian Federation}

\vspace{.5cm}


\vfil

\begin{abstract}

A review of the neutrino conversion and oscillations among the two
neutrino species
(active and
sterile) induced by strong twisting magnetic field is presented and
implications
to neutrinos in neutron star, supernova, the Sun
and interstellar galactic media are discussed. The ``cross-boundary
effect" (CBE)
(i.e., a possible conversion of one half of neutrinos  of  the bunch
from active
into sterile specie) at the surface of neutron star is also
studied for a realistic neutron star structure.

\end{abstract}
\end{center}
\end{titlepage}

\vfill
\vfill
\eject

\section{Introduction}

It is commonly believed that investigations of the neutrino
properties will give a very
important information for better understanding of particle
interactions and for the progress
in theoretical models. One of crucial problems of neutrino physics is
the existance
(or non existance) of the neutrino oscillations.

Neutrino oscillations if observed experimentally would play an
important role in
explanation of different astrophysical phenomena.
The subsequent investigations in this field are
strongly
stimulated first of all by a possible solution of the solar-neutrino
puzzle on the
base of the matter and magnetic field enhancement of spin and flavour
 neutrino
 conversion
 (see, for example, \cite{Wolf}--\cite{Lim} and \cite{Ber}--
\cite{Pal} for a review).
 Another important
motivation
for consideration of neutrino conversion and
oscillations is based on the common belief that these effects may be
involved
in the processes of supernova bursts and cooling of neutron stars
(see \cite{Vol2}--\cite{Ath} and references therein).

The basic idea of the neutrino conversion and oscillations between
the two neutrino
species in vacuum was put forward in \cite{Pon} and  was supplied
\cite{1,Bil} with
the time evolution  analysis of a neutrino beam.

Effects of neutrino interaction with matter of uniform density
on neutrino conversion were considered in \cite{Wolf}.

The flavour conversion in
the case of neutrino propagation in matter with
nonuniform density was studied in \cite{Mikh,Mik} and the resonant
amplification
of neutrino oscillations was predicted  (MSW effect).

The neutrino spin precession in magnetic field  as a possible
solution for the
solar-neutrino problem
was studied  in  \cite{Cis}--\cite{Okun}.
The resonant spin-flavour neutrino conversion that is analogous to
the MSW effect
was applied to the solar neutrino \cite{Akh,Lim} and also to neutrino
from supernova
\cite{Lim}.

It must be mentioned here that
in the most of the performed studies of neutrino
conversion and oscillations between two species in magnetized matter
the
considered strengths of magnetic
field is of the order of $B\leq{10^5 \ G}$ that is quite adequate to
the solar
neutrino problem.
There are also studies and discussions of the neutrino resonant
conversion for
the case of a supernova
accounting for much stronger magnetic fields (see, for example,
\cite{Lim,Vol2,Pelt,Ath}).
However,
in some recent studies in this field the possible influence of
strong magnetic
fields on neutrino conversion and oscillations was not
considered at all
(see, for example, \cite{S-S}--\cite{Raf-S},\cite{APS}).

The magnetic fields of the order of $B\sim 10^{12} - 10^{14} \ G$ are
believed
to exist  at different stages of evolution  of neutron stars.
As an example we mention here new particle interaction phenomena
\cite{ABS, LS} that can be induced by strong magnetic fields and that
can play a visible
role in energetics of neutron stars (see also \cite{R}).
The presence of strong magnetic fields may also influence the
neutrino conversion
and oscillations processes.

 In this paper supposing that
neutrinos have non-vanishing magnetic or/and flavour transition
moments we study the magnetic field induced effects of neutrino
spin and/or
spin-flavour conversion and oscillations between different neutrino
species.  Both the Dirac and Majorana
neutrino conversion and oscillations
effects induced by
strong magnetic fields in the presence of matter, also
accounting for mixing of
neutrinos in vacuum are considered.

We focus on the discussion of the case
when
under the influence of strong enough magnetic
field numerous acts of conversion between the two neutrino specie
occur (i. e.,
neutrino oscillations take place) for each individual neutrino of the
bunch passing
through different  media. In Section 2 a general  analysis of the
problem
is presented and the critical strength of magnetic field $\tilde
B_{cr}$ as
a function of characteristics of  neutrino and matter is introduced
(for magnetic fields $B\geq \tilde B_{cr}$ the magnetic field induced
conversion and
oscillation effects become important). The neutrino oscillations in
magnetic field of a neutron star and the \lq\lq cross-boundary
effect"  (CBE) \cite{St}--\cite{prep}
is discussed (Section 3). The CBE for a realistic neutron star
structure accounting
for variation of the matter density with distance from the centre of
the star is studied
in Section 4. In Section 5 the application of the magnetic field
induced neutrino
oscillations to the supernova reheating problem is discussed, and also
neutrino oscillations in the galactic and twisting solar magnetic
fields are considered.

\vskip5truemm

\section{General Analysis of Neutrino Oscillations
 in Magnetic Field}

The evolution of neutrinos propagating in matter and transverse
twisting
magnetic field $\vec B=\vec B_{\perp}e^{i\phi(t)}$, (the angle
$\phi(t)$
defines the direction of the field in the plane orthogonal to the
neutrino momentum)
is described by the Schr\"odinger-type equation
\begin{equation}
i{d\over d t}\nu(t)={H}\nu(t),
\end{equation}
where the Hamiltonian $H$ can be expressed as a sum of
the four terms (\cite{prep}--\cite{hadr})
\begin{equation}
{H}={H}\sb V+{H}\sb{int}+{H}\sb F+H_{\dot\phi}.
\end{equation}
 Here $H_{V}$ contains a contribution from a vacuum mass matrix,
$H_{int}$
contains a
contribution from neutrino interactions with matter,  $H_{F}$
contains a
contribution from interactions with the magnetic field and
the last term $H_{\dot\phi}$
accounts for the effect of rotation (twisting) of the magnetic field.

If for the case of Dirac neutrinos one uses the bases in which
neutrinos have a definite projection along the direction of
propagation
\begin{equation}
\nu=(\nu_{e_L}, \nu_{{\mu}_L}, \nu_{e_R}, \nu_{{\mu}_R}),
 \end{equation}
then the Hamiltonian is given by (see \cite{Pal},\cite{prep}--
\cite{VW})
\def\afor{$$H\sp D=$$ $$\left(\matrix{-{\Delta
m\sp2\sb{\nu}\over4E\sb{\nu}}c+
V\sp0\sb{\nu\sb e}-{\dot\phi\over2}&{\Delta
m\sp2\sb{\nu}\over4E\sb{\nu}}s&
\mu\sb{ee}B&\mu\sb{e\mu}B\cr
{\Delta m\sp2\sb{\nu}\over4E\sb{\nu}}s&
{\Delta m\sp2\sb{\nu}\over4E\sb{\nu}}c+V\sb{\nu\sb{\mu}}\sp0-
{\dot\phi\over2}&
\mu\sb{\mu e}B&\mu\sb{\mu\mu}B\cr
\mu\sb{ee}B&\mu\sb{\mu e}B&
-{\Delta m\sp2\sb{\nu}\over4E\sb{\nu}}+{\dot\phi\over2}&0\cr
\mu\sb{e\mu}B&\mu\sb{\mu\mu}B&0&
{\Delta
m\sp2\sb{\nu}\over4E\sb{\nu}}+{\dot\phi\over2}\cr}\right).\eqno(3.22a)}

\def\bfor{
\begin{equation}
H\sp D=\left(\matrix{V_{\nu_e}^{-}&{\Delta m^2_{\nu}\over4E\sb{\nu}}s&
\mu\sb{ee}B&\mu\sb{e\mu}B\cr
{\Delta m\sp2\sb{\nu}\over4E\sb{\nu}}s&
V_{\nu_{\mu}}^{-}&
\mu\sb{\mu e}B&\mu\sb{\mu\mu}B\cr
\mu\sb{ee}B&\mu\sb{\mu e}B&
-{\Delta m\sp2\sb{\nu}\over4E\sb{\nu}}+{\dot\phi\over2}&0\cr
\mu\sb{e\mu}B&\mu\sb{\mu\mu}B&0&
{\Delta m\sp2\sb{\nu}\over4E\sb{\nu}}+{\dot\phi\over2}\cr}\right).
\end{equation}
}

\bfor

The Hamiltonian (4) corresponds to the case
of sterile neutrinos $\nu_{e_R}$ and $\nu_{{\mu}_R}$.

For the two Majorana neutrinos in the bases written as
$$\nu=(\nu_{e},\nu_{\mu},\bar\nu_{e},\bar\nu_{\mu})$$
in the corresponding Hamiltonian

\def\onefor{$$H\sp{M}=$$ $$\left(\matrix
{-{\Delta m\sp2\sb{\nu}\over4E\sb{\nu}}c+V\sp0\sb{\nu\sb e}-
{\dot\phi\over2}
&{\Delta m\sp2\sb{\nu}\over4E\sb{\nu}}s&
0&\mu B\cr
{\Delta m\sp2\sb{\nu}\over4E\sb{\nu}}s
&
{\Delta m\sp2\sb{\nu}\over4E\sb{\nu}}c+V\sp0\sb{\nu\sb{\mu}}-
{\dot\phi\over2}
&-\mu B&0\cr
0&-\mu B&
-{\Delta m\sp2\sb{\nu}\over4E\sb{\nu}}c-
V\sp0\sb{\nu\sb e}+{\dot\phi\over2}
&{\Delta m\sp2\sb{\nu}\over4E\sb{\nu}}s\cr
\mu B&0&{\Delta m\sp2\sb{\nu}\over4E\sb{\nu}}s&
{\Delta m\sp2\sb{\nu}\over4E\sb{\nu}}c-
V\sp0\sb{\nu\sb{\mu}}+{\dot\phi\over2}
\cr}\right).$$ $$\eqno(3.22b)$$}

\def\twofor{
\begin{equation}
H^M=\left(\matrix{
V^-_{\nu_e}
&{\Delta m\sp2\sb{\nu}\over4E\sb{\nu}}s&
0&\mu B\cr
{\Delta m\sp2\sb{\nu}\over4E\sb{\nu}}s&
V^-_{\nu_{\mu}}
&-\mu B&0\cr
0&-\mu B&
V_{\nu_e}^+
&{\Delta m\sp2\sb{\nu}\over4E\sb{\nu}}s\cr
\mu B&0&{\Delta m\sp2\sb{\nu}\over4E\sb{\nu}}s&
V_{\nu_{\mu}}^+
\cr}\right),
\end{equation}
}

\def\threefor{$$V^-_{\nu_e}=
-{\Delta m\sp2\sb{\nu}\over4E\sb{\nu}}c+V\sp0\sb{\nu\sb e}-
{\dot\phi\over2},$$
$$
V^-_{\nu_{\mu}}=
{\Delta m\sp2\sb{\nu}\over4E\sb{\nu}}c+V\sp0\sb{\nu\sb{\mu}}-
{\dot\phi\over2},$$
$$V_{\nu_e}^+=
-{\Delta m\sp2\sb{\nu}\over4E\sb{\nu}}c-
V\sp0\sb{\nu\sb e}+{\dot\phi\over2},$$
$$
V_{\nu_{\mu}}^+=
{\Delta m\sp2\sb{\nu}\over4E\sb{\nu}}c-
V\sp0\sb{\nu\sb{\mu}}+{\dot\phi\over2}.$$}
\twofor
where
\threefor
$\mu$ denotes the flavour transition magnetic moment.

Using these Hamiltonians we can consider different neutrino conversion
processes
$\nu_i \rightarrow \nu_j$ and
the corresponding neutrino oscillations $\nu_i \leftrightarrow \nu_j$,
induced by the magnetic field
such as
\begin{equation}
\nu_{e_L} \rightarrow \nu_{e_R}, \
\nu_{e_L} \rightarrow \nu_{{\mu}_R},\
\nu_{e_L} \rightarrow \bar\nu_{{\mu}_R}.\label{proc}
\end{equation}
The probabilities of neutrino conversion from
the
type
$i$ ($\nu_i$) to the type $j$ ($\nu_j$) after  passing a distance $x$
in matter and twisting magnetic field are
\begin{equation}
P(\nu\sb i\to\nu\sb j)=\sin\sp2 2\theta\sb{eff}\sin\sp2\Big({\pi
x\over
L\sb{eff}}
\Big),i\not=j,\label{ver}
\end{equation}
while the survival probabilities are
\begin{equation}
P(\nu\sb i\to\nu\sb i)=1-P(\nu\sb i\to\nu\sb j),
\end{equation}
where the
effective mixing angle $\theta_{eff}$ and effective oscillation length
$L_{eff}$ are given by
\begin{equation}
\tan2\theta\sb{eff}={2\tilde\mu B\over{{\Delta
m_{\nu}^2\over2E}{\A}}-\sqrt{2}
G\sb Fn\sb{eff}+{\dot\phi}},\label{tan}
\end{equation}
\begin{equation}
L\sb{eff}=2\pi\Big[\Big({\Delta m_{\nu}^2\over2E}\A-\sqrt{2}G\sb
Fn\sb{eff}+{\dot\phi}\Big)
\sp2+(2\tilde\mu B)\sp2\Big]\sp{-1/2}.\label{L}
\end{equation}

Note that the effective mixing angle $\theta_{eff}$ and effective
oscillation
length $L_{eff}$ depend on the characteristics of the magnetic field
rotation
$\dot\phi$ along the neutrino pass (see also  \cite{APS,Sm,VW}).

For different neutrino conversion processes (6) $\tilde{\mu}$,
$\A$ and $n_{eff}$ are equal to
\begin{equation}
\tilde\mu=\left\{\matrix{\mu\sb{ee}\hfill&\hfill for \ \nu\sb{e\sb
L}\to
\nu\sb{e\sb R}\cr \mu\sb{e\mu}\hfill&\hfill for\ \nu\sb{e\sb
L}\to\nu\sb{\mu
\sb R}\cr \mu\hfill &\hfill for\ \nu\sb{e\sb L}\to\bar\nu\sb{\mu\sb
R}\cr}
\right.,
\end{equation}
$$ $$

\begin{equation}\A=\left\{\matrix{{1\over2}(\cos2\theta-
1)\hfill&\hfill for\ \nu\sb{e\sb L}
\to\nu\sb{e\sb R}\cr {1\over2}(\cos2\theta+1)\hfill&\hfill for\ \nu
\sb{e\sb L}\to\nu\sb{\mu\sb R}\cr \cos2\theta\hfill &\hfill for\
\nu\sb{e\sb L}
\to\bar\nu\sb{\mu\sb R}\cr}\right.,
 \end{equation}
 $$ $$
\begin{equation}
n\sb{eff}=\left\{\matrix{n\sb e-n\sb n\hfill&\hfill for\ \nu\sb{e\sb
L}
\to\bar\nu\sb{\mu\sb R}\cr n\sb e-{1\over2}n\sb n\hfill&\hfill for\
\nu\sb{e\sb L}
\to\nu\sb{e_R,\mu_R}\cr}\right..\label{n}
\end{equation}

As it was in the case of non-twisting magnetic field  \cite{prep}
the probability (\ref{ver})
  may have a considerable value (the neutrino conversion
processes and oscillations become important) if the following two
 conditions are
valid:

\begin{enumerate}
\item[1)] the \lq\lq amplitude of oscillations"
${\sin^2}2\theta_{eff}$ is
far from zero \hfill\break
(or ${\sin^2}2\theta_{eff} \sim 1$),
\item[ ]and

\item[2)] the length $x$ of the neutrinos path in the medium must be
greater
than the effective oscillation length $L_{eff}$ ($x \sim $ or
$>{L_{eff}
\over 2}$).
\end{enumerate}

The condition 1) is realized if $\tan 2\theta_{eff}$ $\geq 1$, then
from
(9) it follows that at least one of the following two relations must
 be satisfied

 \begin{equation}
{\Delta m_{\nu}^2\over2E}\A-\sqrt{2}G\sb Fn\sb{eff}+\dot\phi=0,\ \
(\tilde\mu B\not=0)
\label{a}
\end{equation}
\begin{equation}
2\tilde\mu B\geq\Big|{\Delta m_{\nu}^2\over2E}\A- \sqrt{2}G\sb
Fn\sb{eff}+\dot\phi\Big|.
\label{b}
\end{equation}

Using the  definitions (see, for example, in\cite{Bahc})
for the oscillation length in vacuum
 $$L_V={4\pi E\over\Delta m_{\nu}^2},
  $$
and  the interaction oscillation length
  $$L_{eff}={2\pi\over\sqrt2G_Fn_{eff}},
  $$
one can write the equations (\ref{tan}), (\ref{L})  as
\def\one{
\begin{equation}tg2\theta\sb{eff}={L\sb VL\sb{int}L_{\dot\phi}\over
L\sb F(\AL\sb{int}\L_{\dot\phi}-L\sb
VL_{\dot\phi}+L_VL_{int})},\end{equation}}
\def\two{\begin{equation}\tan2\theta_{eff}=
L^{-1}_F\Big({\A\over L_V}-{1\over L_{int}}+{1\over
L_{\dot\phi}}\Big)^{-1}
, \end{equation}}
\def\three{\begin{equation}\tan2\theta_{eff}={1/ L_F\over\Big( {\
\A\over L_V} - {1\over L_{int}} + {1\over L_{\dot\phi}}\Big)},
\end{equation}}

\two

\begin{equation}L\sb{eff}=\Big[\Big({\A\over L\sb V}-{1\over
L\sb{int}}+{1\over L_{\dot\phi}}\Big)\sp2+
\Big({1\over L\sb F}\Big)\sp2\Big]\sp{-1/2},\label{Leff}\end{equation}
\def\vst{(E\sb{\nu}=E\sb i\approx\vert\vec p\vert=p)}
where
$$L\sb F={\pi\over\tilde\mu B},\ \
L\sb{\dot\phi}={2\pi\over\dot\phi}.$$

{}From the formula (\ref{ver}) we can obtain the following
expressions for probability in different cases

$$P(\nu\sb i\to\nu\sb j)=$$
\begin{equation}
=\left\{\matrix{\Big({L\sb V\over L\sb F\A}\Big)
\sp2\sin\sp2\Big({\pi x\A\over L\sb V}\Big),\hfill &\hfill {\rm for}
\ L\sb F^{-1}\ll -L\sb{int}^{-1}+L^{-1}_{\dot\phi}\ll
\A L^{-1}\sb V,\cr
\Big({L\sb{int}\over L\sb F}\Big)\sp2\sin\sp2\Big({\pi x\over
L\sb{int}}\Big),
\hfill &\hfill {\rm for} \ L\sb F^{-1}\ll \A L^{-1}\sb V+L^{-
1}_{\dot\phi}\ll L^{-1}\sb{int},\cr
\Big({L_{\dot\phi}\over L_F}\Big)^2\sin^2\Big({\pi x\over
L_{\dot\phi}}\Big),\hfill
&\hfill {\rm for} \ L^{-1}_F\ll \A L_V^{-1}-L_{int}^{-1}\ll L^{-
1}_{\dot\phi},\cr
\sin\sp2\Big({\pi x\over L\sb F}\Big), \hfill &\hfill {\rm for} \ \A
L_V^{-1}-L\sb{int}+L^{-1}_{\dot\phi}=0,
\cr
\to\sin\sp2\Big({\pi x\over L\sb F}\Big),
\hfill &\hfill {\rm for}
\ L\sb F^{-1}\gg \A L_V^{-1}-L_{int}^{-1}+L^{-1}_{\dot\phi} .
\cr  }\right.\end{equation}

The conditions (\ref{a},) (\ref{b}) can be re-written as
\begin{equation}{\A\over L\sb V}-{1\over L\sb{int}}+{1\over
L_{\dot\phi}}=0,\ \
L^{-1}_F\not=0
\end{equation} and
\begin{equation}{1\over L\sb F}\geq\Big\vert{\A\over L\sb V}-{1\over
L\sb{int}}+{1\over L_{\dot\phi}}\Big\vert.\end{equation}

Let us consider the relation (\ref{b}) and suppose that the right-
hand side is not
equal to zero. In  the case of exact equality
   from (\ref{b}) we determine
the critical
strength of magnetic field  \cite{sing}--\cite{hadr}
\begin{equation}\tilde B_{cr}= \left|{1\over2\tilde\mu}\Big({\Delta
m_{\nu}^2 \A\over2E}-
\sqrt{2}G_F n_{eff}+\dot\phi\Big)
\right|\label{Bcr}\end{equation}
that constrain the range ($B \geq\tilde {B}_{cr}$) of field strengths
for which
the value of $\sin^2 2\theta_{eff}$ is not small (i.e., at least is
not less
than $1 \over 2$) for all possible values of the right-hand side term
in (\ref{b}).

It is also possible to express $\tilde B_{cr}$ in a more convenient
for  numerical
estimation form:
\begin{equation}\tilde B\sb{cr}\approx43\Big({\mu_B\over\tilde
\mu}\Big)\Big|-
\Big(2.5{n\sb{eff}\over10^{31} cm^{-3}}\Big)+ \A\Big({\Delta
m\sp2\sb{\nu}\over
eV^2}\Big)\Big({MeV\over E_{\nu}}\Big)+2.5\Big({1m\over
L_{\dot\phi}}\Big)
\Big|\ [Gauss].\label{BCR}\end{equation}

For the case of strong magnetic fields ($B > {\tilde B_{cr}}$),
$\sin^2 2\theta_{eff}
\approx 1$,  we find that for large enough lengths of a neutrino
$\nu_i$
pass given by $x\approx L_{eff}{ k\over 2}, k=1,2,...$ in the
magnetized
medium
characterized by $n_{eff}$ the
probability (\ref{ver}) of conversion process $\nu_i \rightarrow
\nu_j$ can reach the
value of the order of $P(\nu_i \rightarrow \nu_j)\sim1$.

Therefore, the initially
emitted, for example, left-handed neutrino
can undergo
convertion to the right-handed neutrino or to the right-handed
antineutrino
on the path lengths
$x\geq{ L_{eff}\over 2}$.

It is obvious that these oscillation processes take place only in the
presence of
strong magnetic fields $B\gg\tilde B_{cr}$, and the oscillation
length $L_{eff}$,
as it
follows from (\ref{L}), is $L_{eff} \approx L_{F}=\pi/\tilde\mu B$.
For $B\ll\tilde B_{cr}$ the influence of
magnetic field is not important and oscillations (if they exist) are
completely
determined by the vacuum mixing angle and neutrino interaction with
matter.

\vskip5truemm
\section{Neutrino Oscillations in Magnetic Field
 of Neutron Star (Cross-Boundary Effect)}

Now let us consider neutrinos that are produced in the interior of a
neutron
star where magnetic fields of the order of $10^{13} \ G$ (or even a
few orders of
magnitude stronger) can exist (see, for example, \cite{Lan}--
\cite{Lip}). For
definiteness we suppose that initially $\nu_{eL}$'s are produced in
the inner
layers of the
neutron star and  take into account the only one of the conversion
 processes (\ref{proc}),
$\nu_{e_L} \rightarrow \nu_{e_R}$, that can be induced by the
magnetic field on
the neutrino pass from the centre to the surface of the neutron star.

In order to determine the scale of $\tilde B_{cr}$ on the base of
(\ref{Bcr}) and (\ref{BCR})
we use
the following values for characteristics of neutrinos and matter of
the
neutron
star:
$ \mu \sim 10^{-10} \mu_B,\ \mu_B$ is the Bohr magneton, $n_{eff}
\sim 10^{33}\ cm^{-3}$, $\Delta m^2_{\nu}\approx 10^{-4}\ eV^2$,
$\sin 2\theta = 0.1$ and
$E_{\nu} \approx 20\ MeV$. It follows that the main contribution is
given
by the \lq\lq matter" term and for this case
\begin{equation} B_{cr}=1.11\times10^{14}\ G,\end{equation}
(here in contrast with consideration of neutrinos from the Sun
(Section 5)
we do not account for a possible effect of twisting of the magnetic
field).
Magnetic fields just of this order may exist on the surfaces of
neutron stars
\cite{ST,Lip}.

{}From (\ref{L}) for the effective oscillation length we get
$L_{eff}\simeq
1 \ cm$, that is much
less than the characteristic scales of the neutron star structures
(the
thickness
of the crust is, for instance, $L_{crust} \sim 0.1 r_{0}\approx 1\
km$,
$r_0$ is the neutron star radius).

{}From these estimations we can conclude that for neutrinos passing
from the
inner
layers
to the surface the conversion and oscillations effects induced by the
magnetic
field can be important. However, if one is dealing not with a single
neutrino
but
with a bunch of neutrinos that are
emitted in different inner points of the neutron star then the
average of the
$x$
dependent term in formula (\ref{ver})
must be taken. Therefore the probability of $\nu_{eR}$'s appearing
in the initial bunch of $\nu_{e_L}$'s is given by
\begin{equation}\bar P(\nu_{e_R})=
{1 \over 2}
\sin^2 2\theta_{eff}.\end{equation}
It follows that the induced by strong
magnetic
field conversion and oscillations effects could yield in the
approximate equal distribution of neutrinos between
the two neutrino species ( $\sin^2 2\theta_{eff} \sim 1$ if $B\gg
 B_{cr}$); it
also means that there would be a factor of two decrease in amount of
initially emitted $\nu_{eL}$'s in the bunch.

Now let us consider the case of
\lq\lq not too strong
magnetic field", viz., $B<B_{cr}$
along the whole neutrinos path inside the neutron star. If we exclude
the
possibility for the neutrinos to pass through the resonant conversion
point
\cite{Akh,Lim} determined by the Eq.(\ref{a})
 we then get that the neutrino bunch after travelling through the
neutron star will still be composed only of the left-handed neutrinos.
 However,
when the bunch of neutrinos escapes from the neutron star it passes
through a sudden change of density of matter and enters into the
nearly empty
space where $n_{eff}\rightarrow 0$. Effectively it may
result that the neutrinos enter and pass through the region of
strong field ($B>B_{cr}$)
determined on the base of Eq.  (\ref{b}).
The neutrino conversion processes
and oscillations  may thus appear due to the \lq\lq cross-boundary
effect"
(CBE) \cite{sing,prep}.

To consider the CBE we suppose that the magnetic field on the surface
of the
neutron star is of the order of $B \sim B_0=10^{12}\ G$ and that the
strength
 of
the magnetic field decreases with the distance $r$ from the surface
of the neutron star according to the law
\begin{equation}B(r)=B\sb 0\Big({r\sb 0\over
r}\Big)\sp3,\label{Bs}\end{equation}
where $r_0$ is the radius of the neutron star.

The estimation for the
critical field $B^{\prime}_{cr}$ on the base of (\ref{Bcr}) for the
same values
 of
$\mu$, $\Delta
m^2_{\nu}$,
$E_{\nu}$ and $\sin 2\theta$ (again for definiteness the conversion
of the
type
$\nu_{eL} \leftrightarrow \nu_{eR}$ is considered) gives that
\begin{equation}B^{\prime}_{cr}=5.4 \times 10^3\ G.\end{equation}

{}From (\ref{Bs}) it follows that the
magnetic
field exceeds $B^{\prime}_{cr}$ in regions characterized by
\begin{equation}
r\leq
r_{cr} \approx 600 r_0.\end{equation}

   Therefore, along the distances of about $600 r_0$ from the
neutron star the magnetic field exceeds the critical field strength
$B^{\prime}_{cr}$.
{}From the estimation for the effective oscillation length for the
magnetic
field at the surface of the neutron star
\begin{equation}L_{eff}(B \sim B_0) =
{\pi \over {\tilde
\mu B_0}}\simeq10^2{\mu_B\over\tilde\mu }\Big({1G \over B_0}\Big)[m]
=1\ m\end{equation}
it follows that the equal distribution of neutrinos between the
two neutrino
species ($\nu_{e_L}$ and $\nu_{e_R}$) appears after the neutrino bunch
passes through a thin layer $\Delta r \gg 1 \ m$ along which the
decrease of the
magnetic field is still negligible: $\Delta B(\Delta r) \ll B_0$.

Thus, in the case of \lq\lq
not too strong field"
again as it was in the case of
\lq\lq strong field" after the neutrino bunch has passed
a distance $L>L_{eff}$ from the neutron star the equal
distribution of neutrinos among the two species $\nu_{e_L}$ and
$\nu_{e_R}$ appears.

Consider the case when the neutrinos on their path inside the neutron
star
 pass
through the resonant region \cite{Akh,Lim}. In this region the
condition of
Eq.(\ref{a}) is valid. From
(\ref{b}) it follows that for any fixed strength of the magnetic field
there is a layer ( between the two shells with radiuses $r_1$ and
$r_2$)
on the neutrino path to the surface of the neutron star where
effectively the
\lq\lq strong field" case is realized. If the distance $ r_2 - r_1 $
is
greater than
the effective oscillation length $L_{eff} \sim L_{F}$ then after
neutrinos
pass
through this {\it resonant region} again the equal
neutrino distribution between the two neutrino species appears.

Here it is important to note that within the discussed case of the CBE
the adiabatic approximation can be used. In the most general case the
adiabatic
condition is
\begin{equation} (H_{jj}-H_{ii}){\partial\over\partial
r}(H_{ij}+H_{ji})-
   (H_{ij}+H_{ji}){\partial\over\partial r}(H_{jj}-H_{ii})
\ll\hfill
   $$
   $$\hfill2[(H_{jj}-H_{ii})^2+(H_{ij}+H_{ji})^2]^{3/2}
\label{ADIA}              \end{equation}
where $H_{ij}$ are the elements of matrixes of eqs. (4) or (5).
Using expressions for $H_{ij}$ corresponding to the neutrino
conversion
$\nu_i\to$ $\nu_j$  we reduce
the adiabatic condition (29) to the form
\begin{equation} \Big|\tilde B_{cr}{\partial B\over\partial r}-
B{\partial\tilde B_{cr}\over
\partial r}\Big|\ll4\tilde\mu(\tilde
B_{cr}^2+B^2)^{3/2},\label{adia}\end{equation}
that means a slow variation of the magnetic field $B$ and the matter
density $\rho$
$(\tilde B_{cr}=\tilde B_{cr}(n_{eff}))$ with distance $r$. The
magnetic field $B$
is slowly varying function of $r$, whereas $\rho$ undergoes a rather
abrupt change
from the value $\rho_s\sim10^9\ g\times cm^{-3}$ at the surface of
the neutron star
to the value of nearly empty space $\rho_{vac}\to0$. For the chosen
above values of
$\tilde\mu$, $\Delta m_{\nu}^2$, $E_{\nu}$ and $\dot\phi=0$ we get
from (\ref{adia}) that the
adiabatic condition is valid if matter density changes from $\rho_s$
to
$\rho_{vac}$ on the distance $\Delta r\ge\Delta r_{\rho}=10\ cm\ll
L_{eff}$.
Therefore, even in the case of extremely abrupt change of the matter
density
the non-adiabatic effects can be  neglected.

Now in the continuation of discussions of Refs. \cite{St}--\cite{prep}
we should also like to mention
that the CBE can take place not only at the surface of the neutron
star
(when neutrinos escape the matter of the neutron star and start there
path in the empty space where particle number densities $n_e,$ $
n_n,$ $ n_p \to 0$).
The CBE can effectively appear for the Majorana neutrinos passing
through inner layers of the neutron star composed of silicon, oxygen,
nitrogen, carbon and helium. For these shells $n_{eff}=n_e-n_n \to 0$,
that corresponds to isotopically neutral medium (see also \cite{APS}).
Because of the CBE in the inner layers of the neutron star a
reasonable
amount of active neutrinos can be converted to the sterile (non
interacting
with matter) neutrinos that may cause changes in the process of
neutron stars cooling.

In the next Section we shell study the CBE at the surface of the
neutron star
using a model of the neutron star structure provided by a realistic
equation of state
for the matter of neutron star.

\vskip 5true mm

\section{Cross-Boundary Effect for Realistic Neutron Star Structure}

Let us consider the CBE at the surface of the neutron star using a
realistic
model of the star structure to account for change of the matter
density
with distance from the centre of the star.

Neutron star structure  is calculated (see, for example, \cite{ST})
assuming that the equation of state
for neutron star matter, $P=P(\rho)$ ($P$ is the pressure, $\rho$ is
the mass density) at $\rho\geq 2\times10^{14}(g\times cm^{-3})$ is
that of
three-nucleon interaction (TNI) model \cite{FP}. For $\rho$ within
the interval
$4.3\times10^{11} (g\times cm^{-3})
\leq\rho\leq
2\times10^{14}(g\times cm^{-3})$
 the
Baym-Bethe-Pethieck (BBP) equation of state is used. At
$\rho<4.3\times10^{11}(g\times cm^{-3})$  we use the Baym-Pethick-
Sutherland (BPS)
 equation of state \cite{ST, BPS}.

For description of the non-rotating star composed of cold
matter one have to integrate the general-relativistic equation of
hydrostatic balance,
the Tolmen-Oppenheimer-Volkoff (TOV) equation \cite{ST}
\begin{equation} {dP\over dr}=-{G(\rho+P)(m(r)+4{\pi}{r^3}P)
                              \over r^2(1-2Gm(r)/r)} , \label{P}
\end{equation}
\begin{equation}
m(r)=\int\limits_{0}^{(r)}\rho(r^{\prime}){d^3}r^{\prime},
\end{equation}
where $m(r)$ is the mass of the star, $r$ is  radial coordinate
($r=0$ for the centre of the
neutron star), $G$ is the gravitational constant.

In equation (\ref{P}) the preasure $P$ is defined as a function of
the density $\rho$
by equation of state. For outside of neutron star we use the BPS
equation of state.
This model implay that matter is composed of free nuclei, electrons
and neutrons.
The Coulomb  interaction energy is also accounted and the equation of
state can be represented
as a system (see, for example, \cite{ST}):

\begin{equation}\left\{
{\rho=\varepsilon=n_{e}M(A,Z)/Z+{\varepsilon}_{e}
+\varepsilon_L ,\atop
 P=P_e+P_L,} \right. \label{33} \end{equation}
Here
 $\varepsilon$ is the total energy (per unit volume),
 $n_e$ is the electron density, $M(A,Z)$ is the energy
of a nucleus $(A,Z)$,
 $\varepsilon_e$ is the energy of
electrons  without  the energy of the Coulomb
interaction,
$\varepsilon_L$ is the Coulomb interection between electrons and
electrons with
nuclei, $P_e$ is the partial pressure of the electrons and
$P_L={1\over3}\varepsilon_L.$

The values $n_e,\varepsilon_L,P_e,\varepsilon_e$ are represented
as functions of parameter $X_e=p_F^e/m$ ($p_F^e$ is the electron
Fermi momentum) [36]:

\begin{equation} n_{e}={1\over3{\pi}^2{\lambda}_{e}^3}{X}_{e}^3
\end{equation}
(${\lambda}_{e}={1\over{m}_{e}}$ is the electron compton wavelength),
\begin{equation} \varepsilon_L=-1.444{Z}^{2/3}e^2{n}_e^{4/3},
\end{equation}
\begin{equation} P_{e}={m_e\over{\lambda}^3_e}{\Phi}(X_e)
,\end{equation}
where $\Phi(x)={1\over8\pi^2}\{x{(1+x^2)}^{1/2}(1+2x^2)-
\ln[x+(1+{(1+x^2)}^{1/2}]\},$
\begin{equation} \varepsilon_e={m_n\over\lambda^3}\chi(X_e),\label{38}
  \end{equation}
where $\chi(x)={1\over8\pi^2}\{x(1+x^2)^{1/2}(1+2x^2)-
\ln[x+(1+x^2)^{1/2}].$

We assume that the density of the free  neutrons is equal to zero
because it
is below  the density of the neutron drip \cite{ST}.
The values of A and Z used in the sistem (\ref{33})
minimize the energy $\varepsilon$ that corresponds to the equilibrium
nucleus.

For definiteness we  again consider the case of oscillations among
the Majorana neutrinos
($\nu_{e_L}\leftrightarrow\overline{\nu}_{\mu_R}$).
{}From (\ref{n}) whithin the discussed model of the neutron star matter
we have
\begin{equation}
n_{eff}=n_e-\Big({A\over Z}-1\Big)n_e.\label{NEFF}
\end{equation}

Using eqs.
(\ref{33})--(\ref{38}) we perform a computer calculations of
$n_{eff}$ as a function the distance  from the centre of the neutron
star and then determine
$B_{cr}$ (see (\ref{Bcr})) for the different values of $r$.

It is interesting to compare the calculated value of the critical
field $B_{cr}(r)$
and the neutron star magnetic field $B(r)$ for different distances
$r$ from the centre of the neutron star.
We suppose that the neutron star magnetic field for rather large
distances from the surface ($r\geq r_0\approx10\ km$) can be
approximated by eq. (\ref{Bs}).
However, in the close to the star surface layers the magnetic field
may exhibit the more complicated behavior
with variation of $r$.
It is possible to suppose that the field decrease with the
distance from the surface of the star ($r=r_0$) not faster than it
follows from the law $B\sim\rho^{2/3}$ (the field frozen in the
matter).

Using these suggestions on the profile of the magnetic field of the
star we plot
(Fig. 1) the dependence of the critical field $B_{cr}(r)$ (solid line)
and the neutron star magnetic field $B(r)$
on the distance $r$
for the close to the star  surface layers ($r\sim r_0$). It is
supposed that $B(r=r_0)\approx10^{14}G$.
The solid line with dots corresponds to the case when the star
magnetic field
$B\sim\rho^{2/3 }$.

\vfill
\eject

$$ $$
\noindent
\vskip105truemm
\vfill

\noindent
The dashed line shows the behavior of the field for the case when the
field
profile is given by $B(r)\sim1/r^3$.
is

   Fig. 2 shows the dependence of $B_{cr}(r)$
   (solid line)
   and the neutron star magnetic field $B(r)\sim{1/r^3}$
    (solid line with dots)
    on $r$  for  remote distances ($r\geq  r_0$).
   For large distances ($r\gg r_0$)  $B_{cr}(r)$ nearly equal to its
vacuum value
   $$B_{cr}\approx{\Delta m^2_{\nu}\A\over4\tilde\mu E}.$$

{}From these figures it follows that nearly for the whole space from
the surface of the
neutron star ($r_1=10.32\ km$) up to the distances  $r_2\sim10^3\
r_0$ the magnetic field $B(r)$
exceeds the critical field $B_{cr}(r)$. Taking into account that for
the
field $B\sim10^{14}G$ the effective oscillation length $L_{eff}$
determined by eq. (10) is of the order of $\sim1\ cm$ we conclude
that the
magnetic field induced neutrino oscillation effects can be important
for
the space
characterized by $r_1\leq r\leq10^3\ r_0$.

On Fig. 3 we plot the
averaged
probability
$P_{av}=\overline{P}(\overline{\nu}_{\mu_R})$
(similar to one of eq. (24)) that determines the amount of neutrinos
$\overline{\nu}_{\mu_R}$ in the initial bunch of neutrinos
$\nu_{e_L}$ as a function of $r$.
For each individual neutrino (initially emited as $\nu_{e_L}$)
the non-averaged probabilty $P(\overline{\nu}_{\mu_R})$ to detect
the neutrino in the state $\overline{\nu}_{\mu_R}$ oscillate with the
change of $r$ around the average value
$\overline{P}(\overline{\nu}_{\mu_R})$
with the amplitude determined by $B(r)/B_{cr}(r)$.


of the probability variation

$P(\overline{\nu_{\mu_R}})$  in the narrow

\vskip5truemm
\section{Supernova Reheating Problem, Neutrino Oscillations in
Galactic and Twisting Solar Magnetic Field}

The  effects  discussed above of suppression of amount of electron
neutrinos (or
other active neutrinos) induced by strong magnetic fields may have
sufficient
consequences on the reheating phase of a Type II supernova
that can
be used for getting
constraints on the value $\tilde\mu B$. Let us suppose that the
magnetic field
induced neutrino oscillations do not destroy the proposed model
\cite{FMMW}
of about 60 \% \ increase in the supernova
explosion energy. If the magnetic field $B\sim 10^{14}\ G$ exists at
the
 radius of
$r_0= 45 \ km$ from the centre of the hot proto neutron star (the
matter
density
in this region is $\rho\sim 10^{12} \ g/cm^3$) and decreases with
distance
according to (\ref{Bs}) then on the distances $r\sim 160 \ km$ from
the centre the
magnetic field is $\sim 0.6\times 10^{13} \ G$. This field is of
the order of the $ B_{cr}$ determined by (\ref{Bcr}) for the density
$\rho \sim
6\times 10^8 \ g/cm^3$ and the magnetic moment $\tilde\mu\sim 10^{-
10}\mu_B$.
For this case the probability of finding, for example, sterile
$\nu_{eR}$'s
among
the initially emitted
$\nu_{eL}$'s is $\bar P_{\nu_{eL}\rightarrow \nu_{eR}}=0.25$ (the
effective length
(\ref{L}) for this effect is $L_{eff}\sim \ 10\ cm$). Therefore, in
order
to avoid the loss of a substantial amount of energy that will escape
from the
region behind the shock together with the sterile neutrinos, one has
to constrain
the magnetic moment on the level of $\tilde\mu \leq 10^{-11} \mu_B$.

We should like to point out the importance of the resonance
enhancement \cite{Akh,Lim} of neutrino conversion and oscillations
effect in
magnetic fields that may substantially change the energetics of the
shock and
 also give a stringent constraints on the value of $\tilde\mu B$.

It is also interesting to consider the neutrino conversion and
oscillations induced
by the interstellar galactic magnetic fields that are of the
order $B_G\sim 10^{-
6}\ G$. The critical field estimated on the base of eqs. (\ref{Bcr})
for ultra high
energy neutrinos ($E\geq 10^{17}\ eV$) are $\leq 10^{-6}\ G$. Taking
into
account the estimation for the effective oscillation length
$L_{eff}(B\sim
B_G)=10^{20}\ cm$, that is much less than the radius of the galaxy
($R_G\approx
3\times10^{22}\ cm$)
we conclude that in this case the effect of neutrino conversion and
oscillations in
\lq\lq strong magnetic field" can be presented.

Now let us consider the possibility of the twisting magnetic field
induced
neutrino conversion and oscillations (for definiteness we chose the
process
$\nu_{e_ L}\to\nu_{e_R}$) in the convective zone of the Sun.
First of all we estimate the critical field strength, using eqs.
(\ref{Bcr})
with the following values for characteristics of neutrinos and matter:
$\Delta m_{\nu}^2=10^{-4}\ eV^2,\ \sin2\theta=0.1,\ E_{\nu}=20\ MeV,\
n_{eff}\sim n_ e\approx10^{23}\ cm^{-3}$.
For the characteristic of the variation of the field in the
convective zone
along the neutrinos path we suppose that $\dot\phi>0$ and use the
estimation
of Refs.\cite{APS,Sm}:
$L_{\dot\phi}\sim0.1R_{\bigodot}\approx7\times10^7\ m,
\ R_{\bigodot}=7\times10^8\ m$
is the solar radius.
Substituting these values to three terms of eq. (\ref{Bcr}) we get
\begin{equation}\tilde B_{cr}\approx\Big({\mu_B\over
\tilde\mu}\Big)\Big|-10^{-6}-5\times10^{-7}
+1.43\times10^{-6}\Big|\ G=7\times10^{-
8}\Big({\mu_B\over\tilde\mu}\Big)\ G.
\end{equation}

{}From this it is obvious that the account for the twisting of magnetic
field
reduce the value of critical field $\tilde B_{cr}$ to the order of
5\% of
the value $B_{cr}$ that corresponds to the case of non-twisting field.

It is supposed  that the typical value of magnetic fields in the
convective zone
is of the order of $B_{con}\sim10^5\ G$. From (\ref{Bs}) it follows
that $B_{con}$
exceeds the field $\tilde B_{cr}$ if the neutrino magnetic  moment
is greater then $\tilde\mu\geq10^{-12}\mu_B$.

According to the second condition for the magnetic field induced
neutrino
conversion and oscillations become important the effective
oscillation length
$L_{eff}$  have to be of the order or less then the depth of the
convective zone $L_{eff}\leq {1\over2}L_{ cz}$. This last condition
holds if
$\tilde\mu\sim10^{-11}\mu_B$ for the magnetic fields in the
convective zone $B\sim10^5\ G$.

We can conclude that the account for the variation (twisting) of the
magnetic field along the neutrino path in the solar convective zone
relaxes
the critical field strength $\tilde B_{cr}$ to the values which can be
relevant for the stimulation of a visible  neutrino conversion and
oscillations if the neutrino magnetic moment is of the order of
$\tilde\mu\sim10^{-11}\mu_B$.

\vskip5truemm
\section{Summary}
The Dirac and Majorana neutrino conversion and oscillations between
the two
neutrino species induced by  the  magnetic field is considered. We
introduce the critical
magnetic field strength $\tilde B_{cr}(\Delta m^2_{\nu}, \theta,
n_{eff},
E,\tilde\mu,\dot{\phi})$ that determines the range of fields ($B\geq
\tilde B_{cr}$)
for which the magnetic field induced neutrino conversion and
oscillations become important.
This criterion is valid in the case of resonant and non-resonant
amplification of neutrino conversion and oscillations
in magnetic fields.

The criterion is used in the study of the neutrino conversion and
oscillations in magnetic
fields of neutron star, supernova, the Sun and interstellar galactic
media.
The possible conversion of one half of neutrinos from active into
sterile specie
on the neutrino bunch crossing the surface (the ``cross-boundary
effect") of
the neutron star is predicted and discussed in some details.

\vskip5truemm
\section{Acnowledgments}

The authors are thankful to A.Dar, B.Mayer and S.Petcov for useful
discussions.
One of the authors (A.I.S.) should also like to thank
G.Bellittini and M.Greco and all the organizers of the ``Recontres
de Physique de La Thuile" for their kind hospitality.

This work was supported in part  by the Interregional Centre for
Advanced Studies.

\vfill
\eject

\end{document}